\documentclass[12pt]{amsart}
\usepackage[margin=1in]{geometry}

\usepackage{amsmath,amsfonts,amssymb,amsthm}
\usepackage{graphicx}
\usepackage[dvipsnames]{xcolor}
\usepackage{wrapfig}
\usepackage{url}
\usepackage{caption}
\usepackage{subcaption}
\usepackage[justification=centering]{caption}
\usepackage{bbold}
\usepackage{bbm}
\usepackage{hyperref}\hypersetup{colorlinks=True, urlcolor=cyan}
\usepackage{algorithm} 
\usepackage{algpseudocode}

\title[Geospatial bounded confidence with applications to vaccine hesitancy]{A geospatial bounded confidence model including mega-influencers with an application to Covid-19 vaccine hesitancy}
\date{13 October 2022}
\author{Anna Haensch}
\address[Anna Haensch]{Tufts University, Data Intensive Studies Center}
\email[Corresponding author]{anna.haensch@tufts.edu}

\author{Natasa Dragovic}
\address[Natasa Dragovic]{Tufts University, Department of Mathematics}
\email{natasa.dragovic@tufts.edu}

\author{Christoph B\"orgers}
\address[Christoph B\"orgers]{Tufts University, Department of Mathematics}
\email{christoph.borgers@tufts.edu}

\author{Bruce Boghosian}
\address[Bruce Boghosian]{Tufts University, Department of Mathematics}
\email{bruce.boghosian@tufts.edu}

\begin{document}

\maketitle

\begin{abstract}
We introduce a geospatial bounded confidence model with mega-influencers, inspired by Hegselmann and Krause.  The inclusion of geography gives rise to large-scale geospatial patterns evolving out of random initial data; that is, spatial clusters of like-minded agents emerge regardless of initialization. Mega-influencers and stochasticity amplify this effect, and  soften local consensus.  As an application, we consider national views on Covid-19 vaccines.  For a certain set of parameters, our model yields results comparable to real survey results on vaccine hesitancy from late 2020.  

\end{abstract}

\section{Introduction}

Opinions drive human behavior \cite{review_paper}, and opinion formation is a complex multi-scale process, involving characteristics of the individual, local interaction of individuals, social media, mass media etc.  Opinion dynamics have been modeled using approaches inspired by physics \cite{hofstad_networks_book}.  For
surveyes of the 
literature on opinion dynamics see for 
instance \cite{survey_general}, \cite{Proskurnikov_2018}, and \cite{Mastroeni_2019}.

Opinions are formed in part by people talking to their families,  friends,  colleagues,  etc. This is the sort of mechanism that  the bounded confidence model of  \cite{hegsel_krause_2002} aims to capture. It
 is
 just one of several opinion dynamics models that have
appeared in the literature; see for instance  \cite{other_model, other_model2,  gargiulo_2016, other_model3} for others. However, 
our work here starts with the Hegselmann-Krause model.
 
Hegselmann and Krause \cite{hegsel_krause_2015} augmented the model to include the impact of ``radicals" on opinion formation. By their definition, a ``radical" is an individual (or a group of individuals) holding an opinion that is
 extreme (at one end of the opinion spectrum) and unchanging. 
 \cite{mathias_2016} proposed another model including radicals.
 Our Opinion Dynamics Network (ODyN) model includes ``radicals" as well. We call them {\em mega-influencers}, 
 thinking of mass media, prominent
 politicians, etc., and assuming that a mega-influencer is heard by a large
 fraction of the population. 
 
 To a model of {\em opinion space} dynamics 
 with mega-influencers in the style of earlier work such as \cite{hegsel_krause_2015} and \cite{mathias_2016}, we
 add the new feature of {\em geospatial} 
 dynamics. We assume that individuals who are further apart
 from each other
 in two-dimensional space are less likely to influence 
 each others' opinions; 
this is reminiscent of the geometric inhomogeneous random
 graphs of \cite{girgs_algorithm_paper}. The addition
 of a notion of spatial proximity turns out to have
 a very interesting effect: Large-scale geospatial patterns evolve
 out of random initial data. That is, spatial 
 clusters of like-minded agents  (think  ``blue states" and
 ``red states") emerge, regardless of 
 initialization.

Closeness in two-dimensional space can be thought of
 as a stand-in for different notions of closeness; for 
 instance, close family members might be considered ``nearby" even when they live on a different continent. However,  despite the seemingly geography-less nature of the online world, studies have shown \cite{lengyel_2015} that geographic distance remains a key component in the formation and maintenance of social networks. 
 For an extensive review on spatial networks, see \cite{barthelemy_2011}; the networks that we propose here are similar to the hidden variable model for spatial networks presented in Section 3 of \cite{barthelemy_2011}, but futher  include bounded confidence.
 
Finally, we introduce the assumption 
that different people
have different levels of influence, as in a Chung-Lu random graph
 \cite{chung_lu1,chung_lu2,chung_lu3}.  Combing all of these factors, we can compute the probability that two agents speak during a given timestep and update their beliefs accordingly, similar to the random interactions of Weisbuch, Deffuant, et al \cite{weisbuch2002meet}. As a result, opinion clusters no longer become perfectly tight with time, but remain blurred.

Social media does not appear explicitly
in our model.
Social media interactions
can be akin to 
conversations
among friends, family, neighbors, colleagues, in other words the 
sort of interactions modeled by the original 
Hegselmann-Krause model. However, social media users can also be mega-influencers; think of a Twitter account with millions of
followers.

\begin{figure}
    \centering
    \includegraphics[width = .8\textwidth]{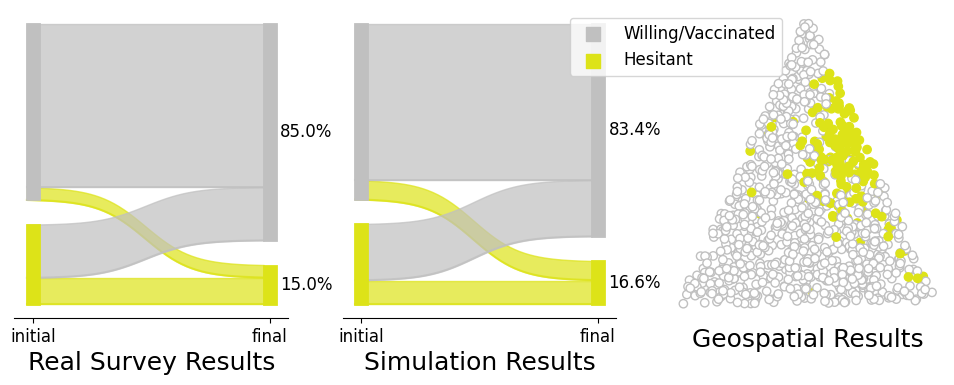}
    \caption{The figure on the far left shows results from a longitudinal survey on Covid-19 vaccine hesitancy \cite{siegler_21}.  The figure in the center shows a similar result from the ODyN model.  The figure on the far right shows the geospatial clustering present in the ODyN model. }
    \label{fig:jama_sample}
\end{figure}

We construct a random graph reflecting all the
features and assumptions discussed above. Vertices represent individuals, and
{\em directed} edges indicate who influences whom. (When individual $v$'s opinion influences that of individual $u$, this does not necessarily imply that $u$ also 
influences $v$.) 

In the simulations presented here, the spatial domain is a triangle, 
and the spatial locations of individuals are independent of each other and uniformly
distributed. The code in the ODyN library 
{\em (Github link removed for anonymity)}
also allows simulations on unions of triangles, where the number of individuals
in each triangle is chosen to be random, Poisson-distributed, with an expected value proportional
to the area of the triangle, possibly with different constants of proportionality for
different triangles. In short, the spatial locations in the ODyN library are
a Poisson point process with a possibly space-dependent rate.

As an example, we consider opinions about Covid-19 vaccination, a topic  of urgent current interest
for which  data are plentiful.  
By April 19, 2021, vaccination was approved for everyone in the US age 16 and older. Despite the fact that the vaccine was free to all residents of the US, many factors impeded widespread vaccination.  There was a lack of availability and access to vaccines in rural areas \cite{murthy_rural}, and vaccine hesitancy was impacted by fear of rare but severe vaccine side-effects \cite{beatty_2021}, social and economic factors \cite{simas_overcoming}, as well as targeted misinformation campaigns and politicization of issues surrounding the vaccine \cite{rabb_plosone}. Others have worked on bounded confidence models and spread of misinformation, for instance see \cite{Douven_2021}. Such models are different from our model, in that they require the presence of a ``ground truth," and therefore there is a concept of mis- and dis-information.  In that and similar works the focus is on how best to organize communities of interacting agents in the presence of a fixed goal informed by the ground truth, and how to achieve this goal most efficiently \cite{Douven_2019},\cite{douven2022art},\cite{Douven_2017}\cite{rosenstock_2017}\cite{zollman_2007}. However, when the question is whether or not to accept a Covid vaccine, there is no objective, unquestionable 
``ground truth."

Since the onset of the Covid-19 pandemic, the extent of hesitancy regarding the vaccine has been tracked at both the national and county levels. For example, the Centers for Disease Control's data portal includes a dataset that provides county level estimates for vaccine hesitancy based on data gathered in the U.S. Census Bureau's Household Pulse Survey,  \cite{cdc_hesitancy_trends}.  Carnegie Mellon University's Delphi Group {\em Covid-19 Trends and Impact Survey} is available through a public API \cite{covidcast} and estimates the extent of vaccine hesitancy at the county level, including changes from week to week.  At the national level, \cite{siegler_21} use survey data to track changes in vaccine hesitancy over time, and notably also include the proportion of people who change their beliefs, becoming either more or less hesitant over time, as shown in the far left of Fig~\ref{fig:jama_sample}. We show how our model can be parameterized to arrive at the empirical results presented in \cite{siegler_21}, and discuss what this might mean in terms of the mechanism of belief proliferation.  Moreover we demonstrate the formation of spatial clusters as seen in the far right of Fig~\ref{fig:jama_sample}. 
Other noteworthy work applying opinion dynamic models to mimic empirical results can be found for example in \cite{Friedkin_2016},\cite{bovet_2020}.

The paper is structured as follows.  We begin by introducing the model, explaining its parameters, dynamics, and  statistics. Next we describe a set of simulations that were carried out to perform model analysis.  Finally, we present the results of these simulations, along with the application to Covid-19 vaccine hesitancy.  All code and data relevant to this paper are distributed in the ODyN library available at {\em (Github link removed for anonymity)}.

\section{Model}\label{model}

\subsection{Directed graph encoding who influences whom}\label{sec:directed_graph}
Let $N$ be a positive integer, and consider $N$ individuals. We use letters such as $u$ and $v$ (for ``vertex"), $1 \leq u, v \leq N$, 
to label individuals. We will construct a random {\em directed} graph in which the individuals are the vertices, with an arrow (a directed edge) from individual $v$ to individual $u$
indicating that $v$ influences the opinion of $u$. We write 
$$
p_{uv} = \mbox{probability of an arrow from $v$ to $u$}.
$$
We do not assume symmetry: $p_{vu}$ need not be equal to $p_{uv}$.\label{paragraph_model1}

\subsection{Spatial locations}\label{sec:spatial_location}
\begin{figure}
    \centering
    \includegraphics[width=.75\linewidth]{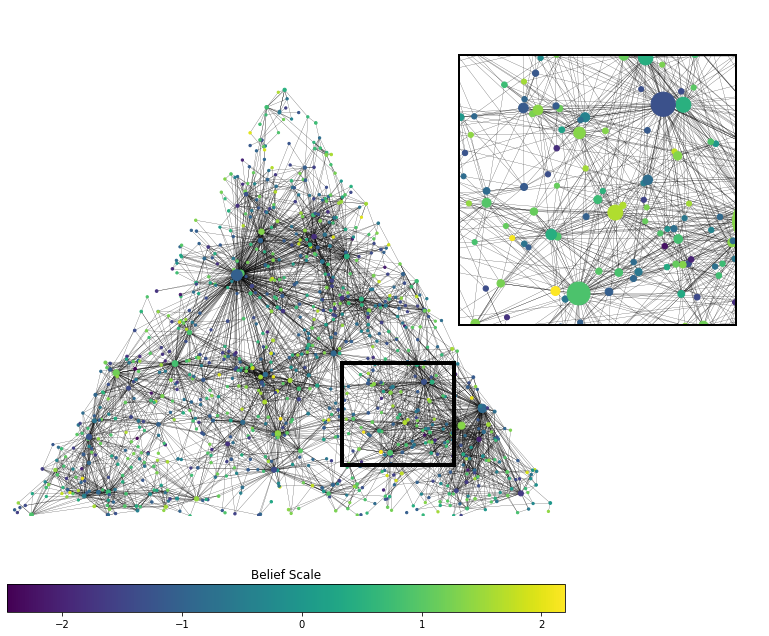}
    \caption{In the triangle above, 1000 agents are placed an an equilateral triangle.  Agents are denoted by circles, where the radius of a circle is a function of the weight of the individual agent and weights are assigned at random with power law exponent 1.5. Agents are randomly assigned beliefs from a 1-dimensional Gaussian mixture model with centers at -1 and 1, both with standard deviation 0.5.  The inset square gives a zoomed in view of one section of the triangle.  Directed edges are then assigned with a probability given by eq.\  \ref{eqn:prob_influence}, although in this image the edges are shown as directionless because of resolution constraints. }\label{fig:triangulation_figure}
    \label{fig:my_label}
\end{figure}

This part of our model is inspired by  \cite{girgs_algorithm_paper}, although several of the details are
different here. 
We assign to individual $v$ a random spatial location $X_v$ in a polygonal domain $D$ in the plane.  In the code available through ODyN, 
$D$ is assumed to be a union of triangles, and the number of individuals per triangle
is taken to be random with Poisson distribution, with a rate that can be different
for different triangles. 
The locations of individuals within 
each triangle are then assumed to be independent and random with uniform distribution (see Fig~\ref{fig:triangulation_figure} for an example). We use triangles because
they are a flexible way of approximating more complicated shapes, and it is straightforward to generate uniformly distributed random points in a triangle.

In the simulations presented here, we simply take
$D$ to be a single triangle, fix $N$, and let the locations $X_1$, $X_2$, $\ldots$, $X_N$
of the individuals be independent, uniformly distributed points in $D$.
We assume
that $p_{uv}$ is a decreasing function of the euclidean distance $\| X_u - X_v \|$.

\subsection{Influence weights} \label{sec:weights}

Following \cite{chung_lu1}, we assign a random {\em influence weight} $W_v > 0$ to each $v$. This weight
determines how likely others are to listen to $v$, not how much weight they assign to $v$'s opinion; the probability $p_{uv}$ is 
an increasing function of $W_v$.

We assume $W_v$ to be a heavy-tailed random variable that is always
greater than 1. Specifically, we assume that for any 
$x>1$, 
\begin{equation}
\label{eq:complementary_distribution}
P(W_v>x) = \frac{1}{x^{\gamma}},
\end{equation}
with some exponent $\gamma>0$. (The parameter $\beta$ of \cite{chung_lu1} is $\gamma+1$.) To generate a random number $W_v$ with the complementary 
distribution function (\ref{eq:complementary_distribution}), draw a uniformly distributed random number $U \in (0,1)$, then 
set 
$$
W_v = U^{-\frac{1}{\gamma}}. 
$$
We will choose a value of $\gamma$ that makes the mean of the distribution of the 
$W_v$ finite: $\gamma>1$. Given this constraint, however, we will choose $\gamma$ to
make the variance of the distribution infinite, so that 
outlying values of $W_v$ become fairly common. The variance is infinite for 
$1 < \gamma \leq 2$, and since within this range, we don't expect the precise value of $\gamma$ to have a qualitative impact on our results, we choose $\gamma = 1.5$.

\subsection{Opinion scores}\label{sec:opinions} Each individual $v$ carries a time-dependent {\em opinion score} $H_v$ between $-1$ and
$1$ in our model, reflecting 
their view on Covid-19 vaccines, ranging from $H_v=-1$ (strong willingness) to $H_v=1$ (strong hesitancy). 
Following \cite{hegsel_krause_2002}
we assume that $p_{uv}=0$ if $|H_u - H_v| \geq b$, where $b>0$ is a threshold. That is, we assume
that $v$ cannot have any impact on $u$'s opinion  if $u$ and $v$ have starkly different views.  Throughout this manuscript, we fix $b=1.5$.  Under this choice of $b$, the classic Hegselmann-Krause model will converge to tight consensus.  However, as we will demonstrate, the ODyN model exhibits other emergent phenomena.

\subsection{Overall formula for the connection probabilities} We define
\begin{eqnarray}\label{eqn:prob_influence}
p_{uv} = \min \left( 1,  \frac{1}{\left( 1 + \| X_u - X_v \|/\lambda \right)^\delta} ~ W_v^\alpha ~ \mathbbm{1}_{|H_u-H_v|<b}  \right)
\end{eqnarray}
where $\mathbbm{1}$ denotes the indicator function. The parameter $\lambda>0$ is a reference
length; we take it to be the diameter of the spatial domain. The parameters
$\alpha>0$ and $\delta>0$ determine the importance of influence weight and spatial
proximity, respectively.

\subsection{Initialization of opinion scores} The influence weights $W_u$ and spatial locations $X_u$ are independent random numbers, 
chosen as outlined above. The opinion scores $H_u$ change with time; see the discussion on Hegselmann-Krause dynamics in
paragraph \ref{paragraph_hk}. We assign a random initial opinion 
score  to each individual, drawn from a Gaussian with standard deviation and mean either $-1$ (with probability $p_{-1}$ or $+1$ (with probability $p_1 = 1 - p_{-1}$).  These assignments 
are made independently of each other, and independently of the $X_u$ and $W_u$. The $p_k$, $k=-1,1$, are chosen 
to reflect publicly available data.

\subsection{Hegselmann-Krause dynamics}
\label{sec:hegselmann_krause}
Denote the opinion scores after $t$ time steps by $H_u(t)$. (We take $t$ to be a non-negative integer here.) Then $H_u(t)$ is the average
of those $H_v(t-1)$ for which either $v=u$, or there is an arrow pointing from $v$ to $u$ at time $t-1$. In words, $u$ 
averages their own opinion with the opinions of those whom $u$ is influenced by. This is  the Hegselmann-Krause model \cite{hegsel_krause_2002}. \label{paragraph_hk}

Since the probabilities $p_{uv}$ depend on $H_u - H_v$, 
they, too, are time-dependent. The connections in the random graph are  re-drawn after each time step, reflecting the fact that people don't necessarily speak and interact with the same people every day.

\subsection{In-degree and clustering coefficient} The in-degree of an individual $u$ is the number of individuals $v$ 
who influence $u$, that is, the number of $v$ for which there is an arrow from $v$ to $u$. We will keep track of the average in-degree.  As the graph is time-dependent, so is the average in-degree. Since every outgoing arrow for one vertex is an incoming arrow for another vertex, the average in-degree equals the average out-degree.

The {\em clustering coefficient} of an individual $u$ is defined as follows. Denote by $k$ the number of 
individuals who influence $u$. 
If $k \leq 1$, then $u$ has clustering coefficient $0$. Otherwise, determine for each of  the $k(k-1)$ ordered pairs of individuals who  influence $u$ 
whether  there is an arrow from the first to the second. The fraction of connected pairs is the clustering
coefficient of $u$.  

\begin{figure}
    \centering
    \includegraphics[width = 4in]{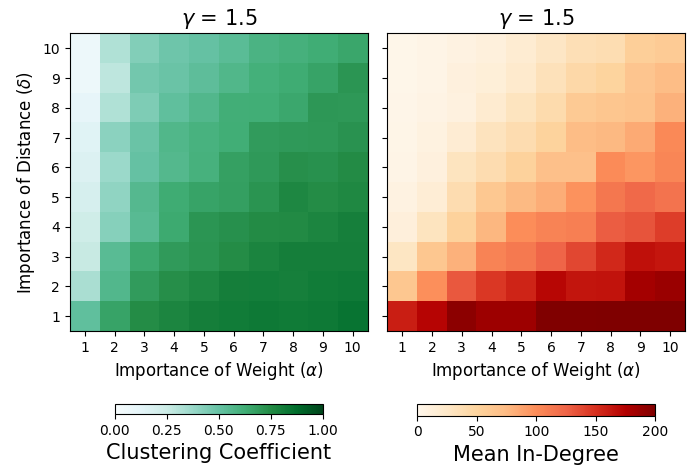}
    \caption{For a model with 1000 agents with $\gamma$ = 1.5 and symmetric initial beliefs centered at -1 and 1 we allow the importance of weight, $\alpha$. and the importance of distance, $\delta$, to vary between 1 and 10.}
    \label{fig:grid_search_alpha_delta_15}
\end{figure}

\begin{figure}
    \centering
    \includegraphics[width=\textwidth]{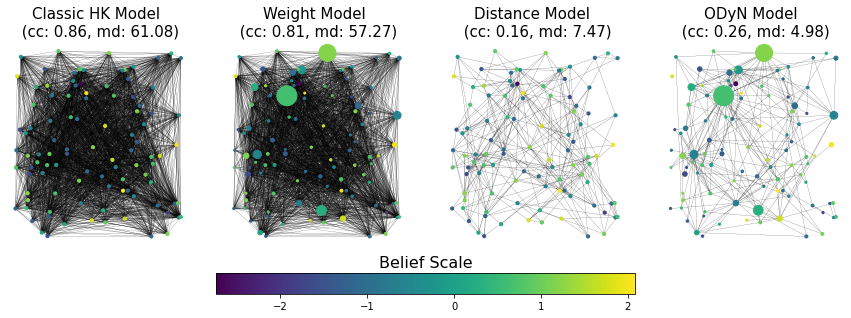}
    \caption{Each of the networks above is initialized with 100 agents holding symmetric beliefs around -1 and 1.  The classic Hegselmann-Krause model (left-most panel) has a very high mean in-degree (denoted md) and clustering coefficient (denoted cc).  Holding all other model parameters fixed, the inclusion of weight (second from left),  distance (second from 
    right) into the full ODyN model (right-most panel) decreases the overall level of connectivity in both the mean in-degree and clustering coefficient. In models for which it is relevant, agents with greater weight are denoted by correspondingly larger dots.}
    \label{fig:ex_neighbor_changing}
\end{figure}

Both the average in-degree and the average clustering coefficient provide a way of evaluating whether our graphs are 
realistic, and therefore help us set parameters.
The average in-degree should not be unrealistically high or low, and the average clustering coefficient should not be 
too low.  To see the overall effect of varying the importance of weight and distance on the clustering coefficients and in-degree, we have performed a grid search across choices of $\alpha$ and $\delta$ with $\gamma = 1.5$, see Fig~\ref{fig:grid_search_alpha_delta_15} (additional results for $\gamma = 1.1$ and $\gamma = 2.0$ can be found in Appendix A,  Fig~\ref{fig:grid_search_alpha_delta_11_20}.  In Fig~\ref{fig:ex_neighbor_changing} we demonstrate the effect of the inclusion of opinions, weights, and distances for a fixed choice of model parameters.

Though a person might interact with a larger number of individuals through their online social networks, or a smaller number of individuals through in-person interactions, surveys have shown that people report feeling genuinely close to between 5 and 10 individuals in their social circle, broadly construed \cite{dunbar}. The clustering coefficient was chosen to be consistent with values for average clustering coefficients on directed graphs using random walks on social networks \cite{hardiman}.

\subsection{Mega-influencers} We add to the model two {\em mega-influencers}, one with opinion score $-1$, referred to as 
the {\em left mega-influencer}, and the other
with opinion score $1$,  the {\em right mega-influencer}. One might think of these as modeling mass media outlets, outspoken governors, etc.  To parallel similar work by Hegselmann and Krause on radicals and charismatic leaders \cite{hegsel_krause_2015}, the mega-influencers hold static beliefs throughout.

The impact of the mega-influencers is modeled as follows. To each individual $u$, we assign two random numbers 
$L_u$ and $R_u$, with
$$
L_u  = \left\{ \begin{array}{cl} 1 & \mbox{with probability $p_L$}, \\
0 & \mbox{otherwise}, 
\end{array}
\right. \hskip 30pt  \mbox{and} \hskip 30pt
R_u  = \left\{ \begin{array}{cl} 1 & \mbox{with probability $p_R$}, \\
0 & \mbox{otherwise}, 
\end{array}
\right. 
$$
where $p_L$ and $p_R$ are further model parameters, with $0 \leq p_L, p_R \leq 1$.
 If $L_u=1$, then $u$ is susceptible to the left mega-influencer. In that case, while $H_u - (-1)< \epsilon$, where $\epsilon>0$ is another model parameter, 
 the opinion score of the left mega-influencer, namely $-1$, 
will be added to the opinions over which $u$ averages in each step of the Hegselmann-Krause dynamics.  Similarly, 
if $R_u=1$, then $u$ is susceptible to the right mega-influencer. In that case, 
while $1 - H_u < \epsilon$,
the opinion score of the right mega-influencer, namely $1$, will be added to the opinions over which 
$u$ averages in each step.
The parameters $b$ and $\epsilon$ play similar roles, for local interactions 
and for mega-influencers, respectively. In the code, they need not be the same, 
but in the simulations presented here, they were the same.

Note that our model assumes that to $u$, mega-influencers do not carry more weight than friends or neighbors. 
The very considerable effect of mega-influencers that we will demonstrate in the computational results is all the more
surprising.

\subsection{Parameterization and model creation} The parameters in our model are $n$, $\lambda$, $\gamma$, $\delta$, $\alpha$, $b$, $\epsilon$,
$p_L$, and $p_R$.  Using the ODyN library, the {\tt OpionionNetworkModel} class can be initialized with these parameters as arguments.  This model can be populated with individuals bearing both weight and belief scores as described in the previous sections using {\tt populate\_model()}.  The belief propagation simulator is loaded as a separate class, {\tt NetworkSimulator,} and network simulations can be carried out on the model with {\tt run\_simulation()}.  Further documentation and demonstrations of this workflow can be found on the project Github page ({\em link removed for anonymity}), and pseudocode for these procedures are given in Algorithms \ref{alg:populate} and \ref{alg:hk} below.\label{paragraph_model17}

\section{Experimental Methodology} \label{methodology}

\subsection{Model Initialization}
The model described in \ref{paragraph_model1} through \ref{paragraph_model17} is generated by Algorithms \ref{alg:populate} and \ref{alg:hk} below.  To populate the network, we generate uniformly distributed random points in a triangle $T$, and as our initial belief distributions, we take symmetric beliefs centered at -1 and 1 with standard deviation 0.5.
We run several experiments varying the reach of mega-influencers from the left and right. Each of the experiments has $n=1000$ agents/vertices and parameters: $\lambda = 1/10$ the diameter of $T$, $\delta = 8$, $\alpha = 2$, $b = 1.5$ and $\epsilon = 1.5$.  These parameters were explicitly chosen to achieve clustering coefficients and in-degrees that were realistic for real-life community interactions, namely, a consistent clustering coefficient of approximately 0.3 as well as an average in-degree around 5.  As noted
earlier, when generating the weights $W_u$, we used $\gamma=1.5$ which yields a heavy-tailed distribution that has finite mean but infinite variance.  Our selection of parameters were chosen to mimic a real-life network in a way that's quantitatively supported by social science research as mentioned earlier. For computational feasibility we restrict our attention to networks with only 1000 nodes, bearing in mind that such networks may be susceptible to edge effects. \label{paragraph_parameter_space}

\begin{algorithm}
\caption{}\label{alg:populate}
\begin{algorithmic}[1]
\Procedure {{\tt populate\_model}}{$n$,$T$, $\gamma$,$(p_0,p_1)$, $\lambda$,$\alpha$, $b$, $\epsilon$, $p_L$, $p_R$}
\State $t_1,t_2,t_3 \gets$ vertices of triangle $T$.
\State ${\tt Agent} \gets $ empty $n\times 2$ position array
\State ${\tt Weight} \gets $ empty $n\times 1$ weight array
\State ${\tt Neighbor} \gets$ $n\times n$ array of zeros.
\State ${\tt MegaInfluences} \gets $ empty $n\times 2$ array.
\For{$i \leq n$}
\State $x,y,w \gets$ sampled from $U(0,1)$
\If{$x+y >1$ is even}
    \State $x \gets 1-x$
    \State $y \gets 1-y$  
\EndIf
\State ${\tt Agent}[i]\gets x \cdot (t_2-t_1) + y \cdot (t_3 - t_1)$
\State ${\tt Weight}[i] \gets $sampled from $U(0,1)$
\EndFor
\State ${\tt Weight} \gets {\tt Weight} ^ {-\frac{1}{\gamma}}$
\State C $\gets$ $n\times 1$ array sampled from $[0,1]$ with probabilities $(p_0,p_1)$, resp.
\State ${\tt Belief[i]} \gets$ sampled from Gaussian centered at C[i] with std. 0.5
\For{$i\leq n$} \label{get_neighbor}
\For{$j\leq n$ with $i\neq j$}
\State $x \gets$ sampled from $U(0,1)$
\If{$x < p_{u_iu_j}$ computed using Eq~\eqref{eqn:prob_influence}}
\State ${\tt Neighbor}[i,j] \gets 1$ 
\EndIf
\EndFor
\EndFor \label{end_get_neighbor}
\State $L \gets$ set of agents with belief within $\epsilon$ of the left influencer.
\State $R \gets$ set of agents with belief within $\epsilon$ of the right influencer. 
\For{$i$ in randomly sampled subset of $L$ with size $p_L \cdot \mid L\mid$}
\State ${\tt MegaInfluencer}[i,0]\gets 1$
\EndFor
\For{$i$ in randomly sampled subset of $R$ with size $p_R \cdot \mid R\mid$}
\State ${\tt MegaInfluencer}[i,1]\gets 1$
\EndFor
\State \textbf{return} ${\tt Agent},{\tt Weight},{\tt Belief}, {\tt Neighbor}, {\tt MegaInfluencer}$ 
\EndProcedure
\end{algorithmic}
\end{algorithm}

With Algorithm \ref{alg:populate} we assign attributes to individual agents such as weight, spatial distance, position in opinion space, and connection to mega-influencers, which we then use to compute the network graph.  Using Algorithm \ref{alg:hk} we synchronously update all opinions.  After each round of opinion updates, the network graph is recomputed holding weight and spatial distance parameters constant, in the manner of adaptive networks \cite{Kozma_2008}.  In the present model, this reflects the fact that somebody who influences my opinion today may not influence it tomorrow, for instance because I may happen not to talk to them tomorrow.

\subsection{Stopping criterion}

For each initialization, we carry out 25 experiments, using Algorithm \ref{alg:hk}, varying the left and right mega-influence (i.e., varying $p_L$ and $p_R$) to the same extent.  A stopping criterion is determined as follows.  At each time step, a 5-time step rolling average in belief change is calculated for each individual.  The community-wide mean of the absolute change in belief is then computed.  When this value drops below .01, the simulation is stopped.  We note that this allows for individuals to have small oscillations in opinion, but overall the community opinion stabilizes.  For brevity, in Algorithm \ref{alg:hk} we indicate this with a Boolean {\tt stopping\_criterion\_satisfied}. This threshold is typically reached in 20 or fewer time steps.\label{stopping_criterion}

\begin{algorithm}
\caption{}\label{alg:hk}
\begin{algorithmic}[1]
\Procedure {{\tt run\_simulation}}{$n$,$\Delta$, $\gamma$, $(p_0,p_1,p_2)$, $\lambda$,$\alpha$, $b$, $\epsilon$, $p_L$, $p_R$}
\State Compute ${\tt Agent},{\tt Weight},{\tt Belief}, {\tt Neighbor}, {\tt MegaInfluencer}$ with Algorithm \ref{alg:populate}.
\While{ {\tt True}}
\For{$j\leq n$}
\State $S\gets \{{\tt Belief}[j]\}\cup \{{\tt Belief}[k] :{\tt Neighbor}[k,j] =1 \text{ for }k\neq j\}$
\If{${\tt MegaInfluencer}[j,0] = 1$} \label{mega_inf} 
\If{${\tt Belief}[j] < \epsilon$}
\State \# Include left mega-influencer belief in $S$.
\State $S\gets S\cup \{\text{left mega-influencer belief (i.e. 0)}\}$
\Else 
\State \# Connect to right mega-influencer with probability $p_R$.
\State $x \gets$ sampled from $U(0,1)$
\If{$x< p_R$}
\State ${\tt MegaInfluencer}[j,0] \gets 0$ and ${\tt MegaInfluencer}[j,1] \gets 1$
\EndIf 
\EndIf
\EndIf \label{end_mega_inf}
\State Repeat steps \ref{mega_inf} - \ref{end_mega_inf} for right mega influencer.
\State \# Propagate all local beliefs. 
\State ${\tt Belief}[j] \gets$ average of beliefs in $S$.
\EndFor
\State \# Recompute network graph using updated beliefs.
\State ${\tt Neighbor} \gets$ Recompute ${\tt Neighbor}$ using steps \ref{get_neighbor} - \ref{end_get_neighbor} in Algorithm \ref{alg:populate}.
\If{{\tt stopping\_criterion\_satisfied}}
\State {\tt break}
\EndIf
\EndWhile
\State \textbf{return} ${\tt Belief}$
\EndProcedure
\end{algorithmic}
\end{algorithm}

\section{Results}\label{results}

\subsection{Model Analysis}
\begin{figure}
    \centering
    \includegraphics[width = .7\textwidth]{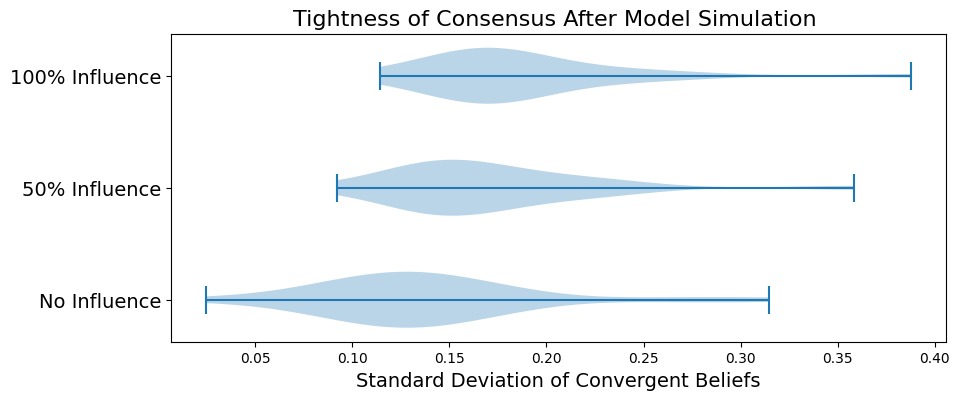}
    \caption{For each setting of mega influence reach (i.e. $p_R$ and $p_L$ equal to 0.0, 0.5 or 1.0) we ran 25 experiments with 1000 agents,  $\alpha = 2$, $b = 1.5$, $\delta = 8$ and $\gamma = 1.5$.  Here we show the distribution in the standard deviation of beliefs at the time of model convergence.}
    \label{fig:violin_plot}
\end{figure}

For each set of parameters, we ran 25 random seeded simulations.  In Fig~\ref{fig:violin_plot} we show how the variance of beliefs when the stopping criterion is met (see \ref{stopping_criterion}) is distributed for each set of 25 experiments. From this plot we can clearly see that a loose consensus is reached in every scenario, but always with a significantly higher variance than what is seen in a classical Hegselmann-Krause model.  In the absence of mega-influencers the standard deviation of beliefs at the final time is typically around 0.13.  Given 50\% mega-influencer reach it is most often near 0.15 and for 100\% mega-influencer reach it is near 0.18.

\begin{figure}
    \centering
    \includegraphics[width = .8\textwidth]{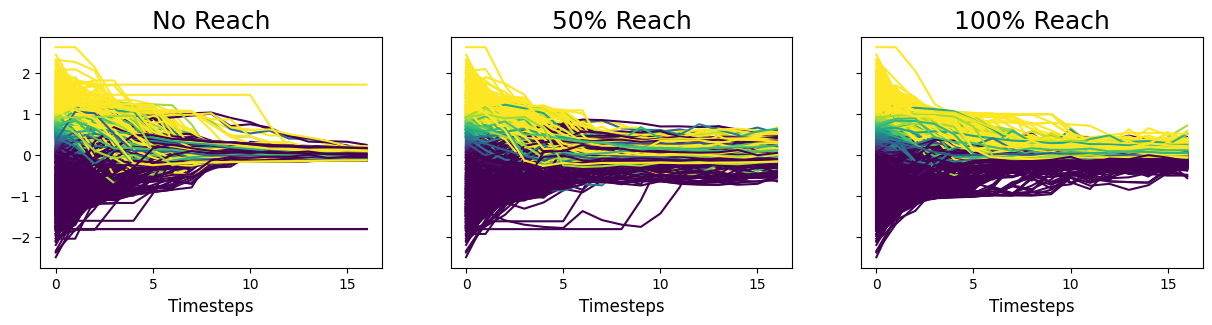}
    \caption{Sample simulation results for each setting of mega influence reach (i.e. $p_R$ and $p_L$ equal to 0.0, 0.5 or 1.0) with 1000 agents,  $\alpha = 2$, $b = 1.5$, $\delta = 8$ and $\gamma = 1.5$.  We see a wide spread of final beliefs as well as holdout individuals.}
    \label{fig:mega_influencer_reach}
\end{figure}

As further evidence of this behavior, we show simulation plots for one distinct iteration in Fig~\ref{fig:mega_influencer_reach}.  The simulation here has reached a stopping point after 16 time steps.  We observe that even in the absence of mega-influencers, agents who are situated geographically far from other agents with similar beliefs can still end up stuck in their initial beliefs and therefore become holdouts.  

\begin{figure}
    \centering
    \includegraphics[width = \textwidth]{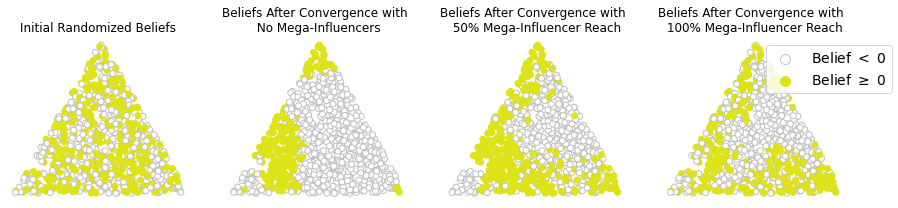}
    \caption{Sample geospatial simulation results for each setting of mega influence reach (i.e. $p_R$ and $p_L$ equal to 0.0, 0.5 or 1.0) with 1000 agents,  $\alpha = 2$, $b = 1.5$, $\delta = 8$ and $\gamma = 1.5$. Agents are shaded according to belief.}
    \label{fig:geospatial_plot}
\end{figure}

In Fig~\ref{fig:geospatial_plot} we take a more geospatial view of the same model, and look at the beliefs as situated in space.  For simplicity we color agents white if their beliefs are less than 0 and yellow if their beliefs are greater than or equal to 0.  Although the relative magnitudes in each direction aren't shown, we can confirm from Fig~\ref{fig:mega_influencer_reach} that there is indeed a broad spread of beliefs away from 0, especially in the presence of mega-influencers.  We observe that despite randomly initialized beliefs, spatially close communities of like-minded agents eventually emerge.  This is seen as patchy yellow neighborhoods in Fig.~\ref{fig:geospatial_plot}.

\subsection{Application to Vaccine Hesitancy}

\cite{siegler_21} present a cohort study of changes in vaccine hesitancy from a baseline survey taken between August 9 and December 8, 2020 and a follow-up survey taken March 2 to April 21, 2021. At baseline, 69\% of respondents indicate that they are “likely” or “very likely” to receive the vaccine and accordingly, are classified as {\em willing}, while the remaining 31\% indicate that they are “very unlikely,” “unlikely,” or “unsure” about receiving the vaccine and are therefore classified as {\em hesitant}.  At the time of the follow-up survey, 47\% of respondents have been vaccinated, 38\% are willing to receive the vaccine but hadn't yet done so, and  15\%  were hesitant.  Notably, there were individuals from both initial cohorts that fell among the vaccinated, willing, and hesitant cohorts at follow-up, that is, people changed their minds to become both more willing and less willing over time.  The complete data from this study can be found in a table in \cite{siegler_21}.

\begin{figure}
    \centering
    \includegraphics[width = .7\textwidth]{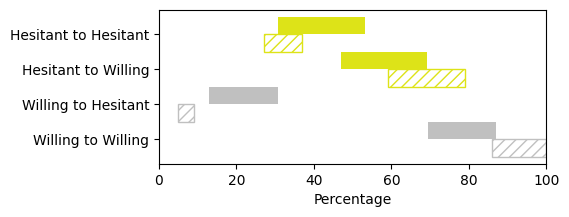}
    \caption{The 95\% confidence intervals from \cite{siegler_21} are shown as cross-hatched bars, and the 95\% confidence intervals from 30 simulations carried out with the ODyN model ar shown as solid bars.  Bars are are futher color-coded to denote whether the cohort began as Willing (i.e. Belief <0) or Hesitant (i.e. Belief $\geq$ 0).  For example in the case of ``Willing to Willing" our model suggests that 69\% to 87\% of the initially willing remained willing, whereas data from \cite{siegler_21} suggests that 86\% to 100\% of the initally willing remained willing.}  
    \label{fig:jama_comparison}
\end{figure}

Using the ODyN model, we are able to recreate these results.  We seed 
the model with 1000 agents and initial beliefs centered at -1 and 1 
with probabilities .69 and .31 respectively, and standard deviation 0.5.  Fixing model parameters $\alpha = 2$, $b= 1.5$, $\delta = 8$ we perform a grid search across different choices for left and right mega-influencer reach.  The choice of parameters which best reproduced the real survey results was a left influencer reach of 35\% and right influencer reach of 75\%.  These results are shown in Fig.~\ref{fig:jama_sample}. For other values of $p_L$ and $p_R$, results are shown in Fig~\ref{fig:jama_model} of the Appendix.  The plots shown in Fig.~\ref{fig:jama_sample} and \ref{fig:jama_model} are for individual simulations, but in Fig.~\ref{fig:jama_comparison} we demonstrate how our model compares with \cite{siegler_21} across multliple simulations.

\section{Conclusion} \label{conclusion}

The most interesting outcome of our model is the geospatial clustering.  In particular, even an initially randomly mixed population evolves into patches of geospatially consistent beliefs.  This suggests that even if there were no ``blue states" and ``red states" they would eventually emerge for mathematical reasons. Even individuals who are initially beyond the bounded confidence thresholds of their neighbors will eventually have their views softened.

Our  model also shows the spread of opinions observed with the introduction of mega-influencers.  Unlike in previous models including radicals with static beliefs, such as in \cite{hegsel_krause_2015}, the combination of stochasticity and mega-influencers in our model has the effect that opinions eventually stabilize but never tightly around a single or double consensus.  Mega-influencers result in a more diffuse set of individual beliefs.

We demonstrate how a certain set of model parameters recreate real life data related to changing opinions around vaccine hesitancy. As in real life, our model shows agents changing their minds to become both more accepting and more hesitant.  Of note is the fact that a substantial influence from the right is needed to prevent public opinion from almost entirely shifting towards vaccine acceptance.  This is shown convincingly in Fig.~\ref{fig:jama_model} where the presence of any reach from the left is enough to overpower all but the most broadly reaching right influencers.

Our geospatial distance can be interpreted as geographic, or some other notion of distance. Physical proximity is an important component of political and ideological opinion formation.  On the other hand, social media makes spatial proximity less important,  but might tend to make people interact more selectively with the like-minded as both a consequence of social and algorithmic behavioral drivers \cite{cinelli_social}, although this is a point of discussion \cite{Bakshy_2015}. One could attempt to model this effect by changing parameters in our model, making spatial proximity less important, and making like-mindedness more important, in determining the probabilities $p_{uv}$. It might also be worthwhile to consider the role of bots in networks and opinion formation online as in \cite{Keijzer_2021}.

Our results also suggest future work on the dependence on the parameters 
$b$ and $\epsilon$. We intend to work on scalable sampling algorithms for combining triangles into other more complex geometries. We plan to attempt to understand how time steps in our model map onto real time. Another feature to be added to the model in the future would be the effects of central interventions such as government or workplace vaccine mandates in the case of Covid-19 vaccinations. The connection between beliefs and the spread of misinformation could also be incorporated into our model. 

\section{Appendix A: Additional Supporting Figures}\label{appendix}

\begin{figure}[h]
    \centering
    \includegraphics[width = 4in]{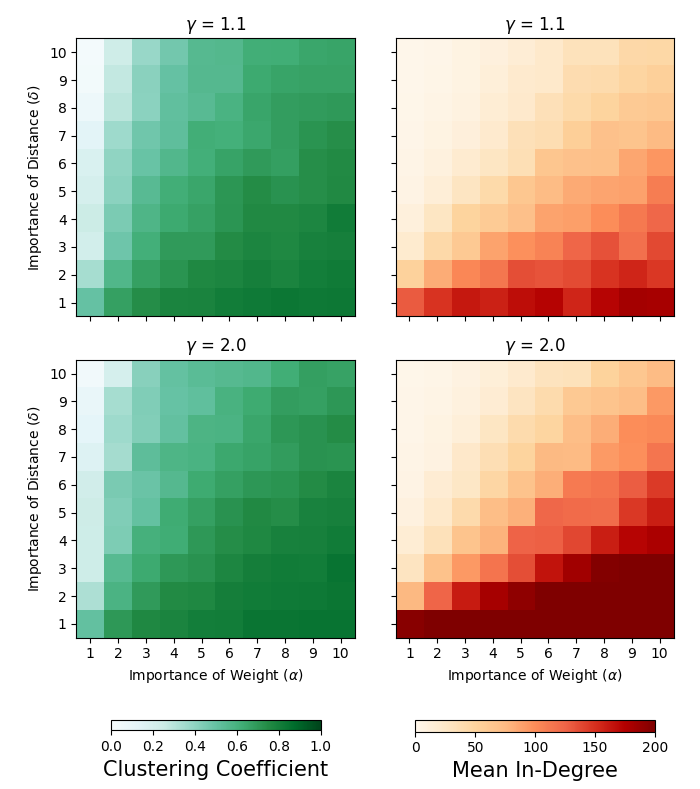}
    \caption{For a model with 1000 agents with and symmetric initial beliefs centered at -1 and 1 we allow the importance of weight, $\alpha$. and the importance of distance, $\delta$, to vary between 1 and 10 with $\gamma = 1.1$ (top) and $\gamma = 2.0$ (bottom).}
    \label{fig:grid_search_alpha_delta_11_20}
\end{figure}

\begin{figure}
    \centering
    \includegraphics[width = \textwidth]{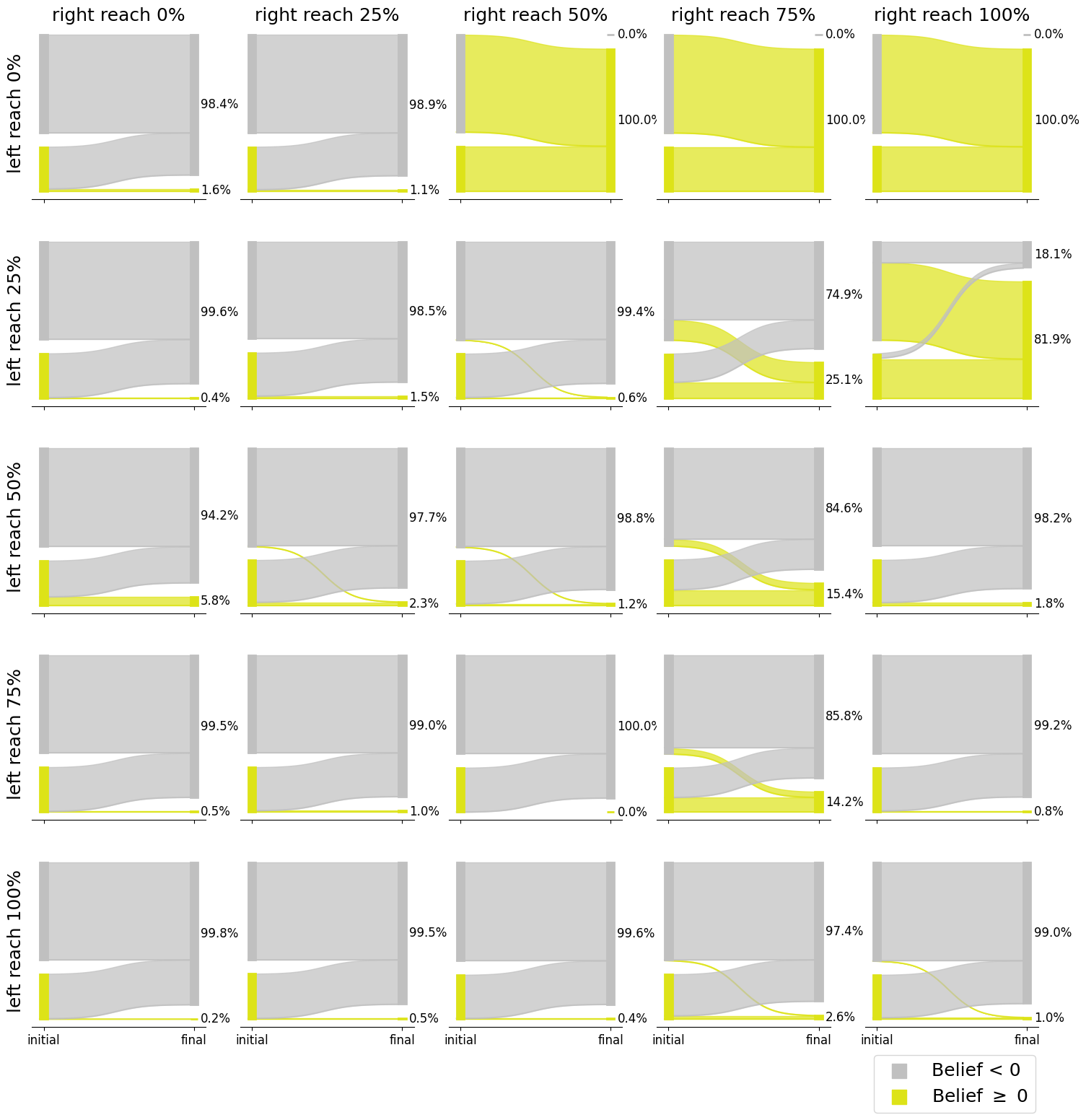}
    \caption{The model is seeded with 1000 agents, initial beliefs centered at -1 and 1 with probabilities .69 and .31, respectively, $\alpha$= 2, $b$ = 1.5, $\delta = 8$.  The alluvial plot shows the overall cohort changes in belief from model initialization to the model convergence.}
    \label{fig:jama_model}
\end{figure}

\bibliographystyle{plain}
\bibliography{my_bib.bib}

\end{document}